\documentclass[namedreferences]{solarphysics}

\usepackage[hyperref,optionalrh]{spr-sola-addons} % For Solar Physics 
\usepackage{graphicx}        % For eps figures, newer & more powerfull
\usepackage{color}           % For color text: \color command
\chardef\us=`\_

\begin{document}

\begin{article}
\begin{opening}

\title{Time path of turbulence and multi-fractality of magnetic field in the evolution of an active region   }

\author[addressref=aff1,corref,email={vabramenko@gmail.com}]{\inits{V.}\fnm{Valentina}~\lnm{Abramenko}}

\address[id=aff1]{Crimean Astrophysical Observatory of Russian Academy of Sciences, Nauchny 298409,
	Bakhchisaray, Republic of Crimea}

\runningauthor{V. Abramenko }
\runningtitle{\textit{Solar Physics} Time path of turbulence in an active region }

\begin{abstract}
Magnetograms acquired with the Helioseismic and Magnetic Imager (HMI) on board the Solar Dynamics Observatory (SDO) were used to calculate and analyze time variations of turbulence and multifractality in the photosphere during the development and flaring of a mature active region NOAA 13354 during its passage across the solar disk. Turbulence was explored with 2D magnetic power spectra from magnetograms, and multifractality was analyzed using the structure functions of magnetograms. 
 Time variations of the magnetic power spectrum exponent $\alpha$ and of the multifractalty exponent $\kappa$ demonstrate no pre-flare or post-flare abrupt peculiarities, instead, 
long periods of stability with smooth transitions into other conditions were observed. 
A conclusion was inferred that the turbulence and multifractality time path in the photospheric magnetic field does not follow the timing of single flares,  however, it tends to correspond to the levels of the magneto-morphological complexity and flaring productivity of an AR. So, in the sense of self-organized criticality (SOC), the photosphere, being in the state of self-organization, evolves independently from the highly intermittent, SOC-state corona.

\end{abstract}
\end{opening}

\section{Introduction}
\label{S-Introduction} 
The solar surface is thought to be a thin layer of several hundred kilometers, hosting sunspots (sunspot groups), which are visible as dark features on white-light images and surviving usually several weeks. The above atmosphere is spread over a huge volume and undergoes fast, sporadic changes continuously, especially above sunspots. In the photosphere of the sunspot group, on one hand, and in the atmosphere above, on the other hand, we observe different physical conditions (temperature, gas pressure, velocities, etc.), different temporal rhythm of changes, different spatial scales of phenomena. Albeit, all this makes the joined system  - a solar active region (AR). How the connection between the photosphere and corona works in an AR  - a question which is still open. 
What kind of phenomena and on what temporal and spatial scales  provide an energy and/or information exchange between the slow-varying photosphere and the bursty developing corona?  
The most plausible agent to connect the photosphere with upper atmospheric layers is the magnetic field: the magnetic field lines are thought to penetrate from the solar interior through the photosphere into the corona and eventually return back to the Sun. Moreover, it is thought than the magnetic field is capable to deliver into the corona the free magnetic energy necessary to feed the solar flaring process \citep{Fang2012,Vekstein2016} (see, however, a recent publication \citet{Karimov2024}).  

Concerning the flaring process, what do we know about the flare-related interaction between the photosphere and corona from observations? The so-called white-light flares \citep{Angle1961,Hudson1972,1975SoPh...42..421C,2000A&A...354..691G,Hudson2016} are a response of the photosphere to strong beams of charged particles penetrating deep down during the flaring process occurring in the upper layers. Sunquakes \citep{Donea2011,2014ApJ...796...85J,2016ApJ...831...42R} also seem to be the photospheric response to flares via seismic waves. Investigations of the flare-related magnetic field changes in the photosphere are rather scanty.   \citet{Chen2007} and \citet{Wang2007} revealed that during a flare, the magnetic field in a sunspot can become more vertical, which can be explained by reconnection of magnetic field lines in the corona during the flare.  \citet{Burtseva2015} showed that strongest flare-related changes in the photospheric magnetic field occur in pixels, where the kernels of the hard X-ray emission are located; the peak in the magnetic changes can occur before or after the X-ray peak. These authors interpret the result as a response of the photosphere to the current sheet formation and subsequent reconnection in the corona.   

A question about the reverse connection - from the photosphere to corona  - seems to be much more complicated and less explored. The hypothetical concept that the observed growth of the sunspot area is accompanied by the emergence of magnetic flux tubes seems to be supported by numerical simulations \citep{Abbett2000,Archontis2008,Fang2010} (see also a review by \citet{Fan2009}), however, cannot help much in understanding the observed burst-like behavior of coronal magnetic energy release. The energy transfer mechanisms via magneto-hydrodynamic and acoustic waves for coronal heating \citep{Bingham2010,Verdini2012} and solar wind acceleration \citep{Ofman2010} are widely explored,  however, these mechanisms can hardly help much in explaining  the spontaneous flaring process, covering a huge variety of spatial and temporal scales. Again, we return to the magnetic field that penetrates and consolidates an active region, from sub-photosphere to the upper layers of the corona. As it is mentioned above, the energy necessary for flaring comes from the photosphere to the corona as energy of twisted magnetic field. How does the mechanism work? Presumably, twisting of the magnetic field lines and electric current generation due to turbulent motions of solar plasma provides the magnetic energy for flaring (see an excellent review of observational aspects in \citet{Archontis2008}). 

To this end, a measure of complexity of the photospheric field should be in accordance with the flaring productivity of an AR. 

 The key question is what an entity/parameter should be taken as a measure of complexity. The first thing that comes to mind is the parameters of turbulence and multi-fractality. 
There were several publications demonstrating this concept from an observational point of view. Thus, 
in \citet{Abramenko2005PS} based on high-resolution MDI magnetograms for 16 ARs and quantitative index of flare activity (derived  as a sum of GOES flare magnitudes over the interval of observations of an AR), it was reported that the magnetic energy spectrum tends to be steeper for ARs with higher flare index.  
\citet{Hewett2008,McAteer2010}, using the full-disk and high-resolution MDI data for ARs NOAA 9077, 10488 explored the spectra of multi-fractatity and magnetic energy and concluded that these measures are promising to "... predict the true flare potential of individual active region in near-realtime."  \citet{Conlon2010} explored 5 ARs using a sophisticated wavelet transform method to calculate a fractal dimension and a Holder exponent of the magnetic field. Authors reported that "when sufficient flux is present in an active region for a period of time, it must be structured with a fractal dimension greater than 1.2, and a Holder exponent greater than -0.7, in order to produce M and X class flares." 

In  \citet{Abramenko2010PS,Abramenko2010Int} a statistical study was performed for 217 ARs on the basis of high-resolution MDI magnetograms: a flare index was plotted versus the magnetic power spectrum index, $\alpha$ and the multi-fractality index, $\kappa$, with correlation coefficients of 0.53 and 0.63, respectively. The  scattering in both diagrams was rather broad (see Fig. 2a in \citet{Abramenko2010PS} and Fig. 7 in \citet{Abramenko2010Int}), so that the rather high correlation was ensured by a small number of highly-flaring ARs. These authors emphasized that application of their methods (Fourier spectrum and structure functions scaling) requires high-resolution data (see more details below). These results suggested that there exists a statistical tendency of active regions to show stronger degree of turbulence and multifractality with the enhancement of flaring capability.  

Using the same methods, augmented by the box-counting routine, \citet{Georgoulis2012} (the paper was accepted for publication in 2010) analyzed the 4-hour time series of low-resolution MDI full-disk magnetograms for 370 ARs, predominantly segregating them into two categories: flaring (with M- and X-class flares) and non-flaring (the rest). Author examined the fractal dimension, a multifractality-related exponent and the magnetic power spectrum exponent, comparing their pre-flare and peak values in flaring regions with the peak values in non-flaring regions and found that none of these parameters could reliably distinguish the flaring ARs from the non-flaring ones. In spite of this rather questionable statistical inference, the paper by  \citet{Georgoulis2012} was very useful in terms of stimulation to search for new parameters and approaches for flare forecasting, in particular, the role of magneto-morphological properties of an active region was pointed out for the first time in this paper.

Now, we should take a step away from the flare prediction problem and to concentrate on the more intrinsic questions: How is the turbulence/fractality in the photosphere related to the SOC-state in corona? 

How turbulence and multifractality behave in an individual AR while a flare is being prepared and going on in the corona? An analysis of temporal variations of turbulence and multifractality of a suitable AR can help to clarify the question. The AR must be observed  from the appearance on the disk for several days while it passes across the disk, must produce a strong flare when located near the central meridian, must display some substantial  restructuring (say, an emergence of a $\delta$-structure). An analysis of such AR was the goal of this study.

 \section{Data}
 \label{S-Data} 
\begin{figure}
\centerline{\includegraphics[width=1.0\textwidth,clip=]{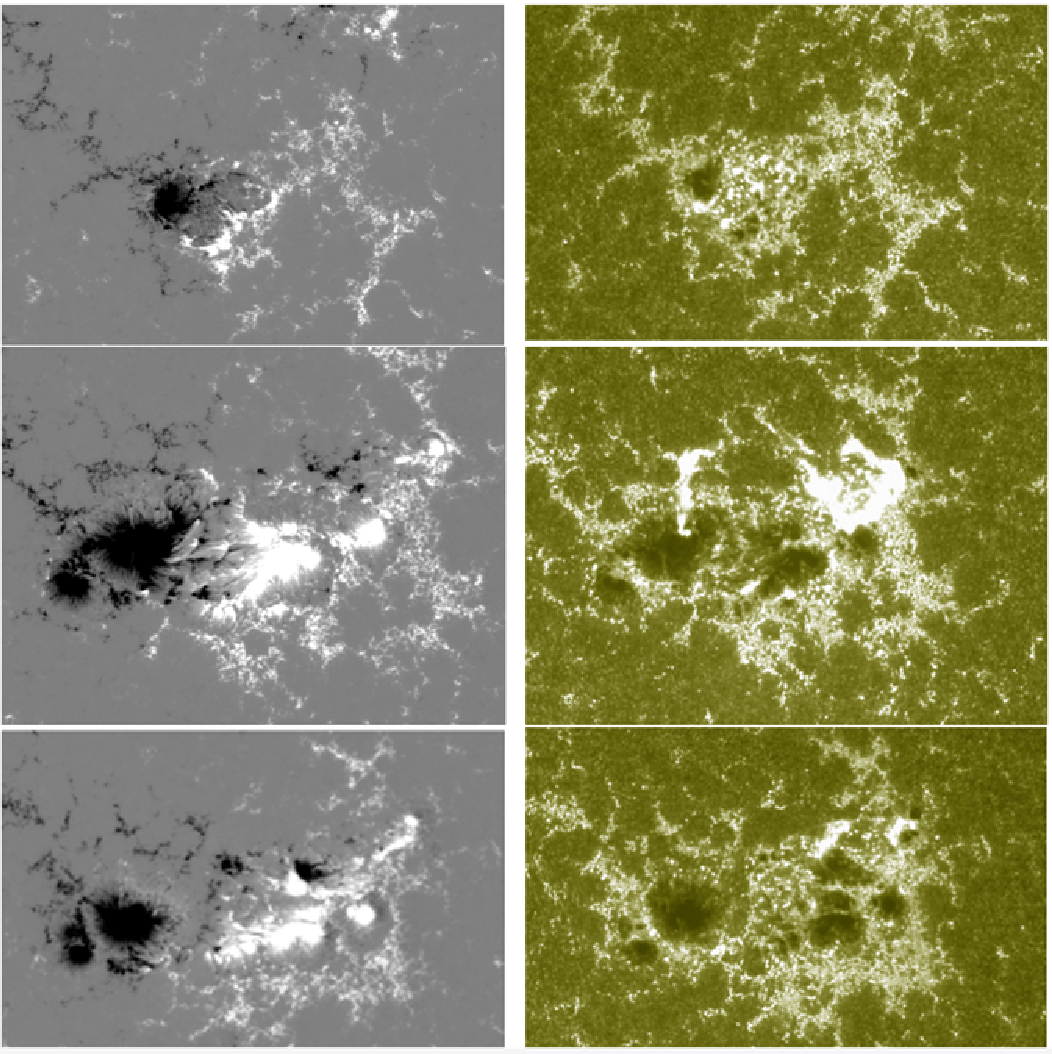}}
\caption{ Left column - SDO/hmi.sharp-cea-720s magnetograms of AR NOAA 13354. All magnetograms are of the same spatial scale, and the horizontal dimension is 260 Mm (720 pixels). The magnetograms are scaled from -800 Mx$\cdot$cm$^{-2}$ (black) to 800 Mx$\cdot$cm$^{-2}$ (white). East is to the left, north is to the top. Right column - co-temporal images in the AIA 1600~\r{A} spectral line. Data are taken at:  27 June 2023 (00:36 UT) - top row, 29 June 2023 (magnetogram: 14:12 UT, AIA: 14:15 UT) - middle row, and 30 July 2023 (19:48 UT) - bottom row. The middle-row data reflect the maximum of the M3.8 flare. The bottom-row data show a typical situation during the development of the newly emerged $\delta$-structure.}
\label{fig1}
\end{figure}

 The choice of an active region which would satisfy the above requirements was not an easy issue. Finally, we selected AR NOAA 13354 (Figure \ref{fig1}), which appeared on the eastern part of the solar disk on 26.06.2023. The emergence was very fast: by the end of 28.06.2023, when it was on the central meridian, the AR became a large, well-developed multipolar structure with the total unsigned flux of 4.2 $\cdot$ 10$^{22}$ Mx. The AR was practically isolated: no other significant active regions on the disk were noticeable. We monitored the AR for 5 days: from 27.06.2023/00:00 UT till 01.07.2023/23:48 UT. 
  The AR was located at N15 E25 in the beginning of our monitoring, and by the end of the monitoring it was rotated to W50. Further monitoring was useless due to the projection effect. (Note that here I avoided usage of the transverse magnetic field data.) 
 Approximately at noon of 30.06.2023 a moderate-size $\delta$-structure starts to form in the leading part of the AR (see the left bottom frame in Figure \ref{fig1}). During the interval of our analysis,  the AR launched several weak C-class flares and only one moderate M3.8-class flare on 29.06.2023/14:00 UT when the AR was located near the central meridian, approximately at W12 (see the timing of flaring in Figures \ref{fig3}, \ref{fig5}, red lines). 
  Later on 02.07.2023 at 22:54 UT, the AR launched the X1.1 flare.

 Flare productivity of the AR was measured by the flare index, $FI$, introduced
 in \citet{Abramenko2005PS} as
 \begin{equation}
 	FI=(100 S^{(X)}+10 S^{(M)}+ S^{(C)}+0.1 S^{(B)})/\tau.
 	\label{FI}
 \end{equation}
 Here, $S^{(j)}$ is the sum of all GOES flare magnitudes in the $j$-th X-ray
 class:
 \begin{equation}
 	S^{(j)}=\sum_{i=1}^{N_j} I_i^{(j)},
 	\label{AA}
 \end{equation}
 where $N_j=N_X, N_M, N_C$ and $N_B$ are the numbers of flares of X, M, C and B
 classes, respectively, that occurred in a given active region during the considered time interval $\tau$ (in days).
 $I_i^{(j)}=I_i^{(X)}, I_i^{(M)}, I_i^{(C)}$ and $I_i^{(B)}$ are GOES
 magnitudes of X, M, C and B flares. 
 Usually, the interval $\tau$ was taken as the entire interval $\tau(total)$ of the AR's presence on the disk.  For the AR 13354 $\tau(total)$ = 8.5 days, which gives us $FI(total)$=35. To follow the dynamics of flaring activity, it was interesting to explore $FI$ for smaller $\tau$. A 4-day window was shifted by 1 day, so that $FI(4d)$-values of 25, 23, 56, 44, 47 were calculated for intervals centered at 29.06/00:00 UT, 30.06/00:00 UT, 01.07/00:00 UT, 02.07/00:00 UT, and 03.06/00:00 UT. The data are plotted in Figures \ref{fig3} and \ref{fig5} with blue stars. So, we observe that the flare index grows as the AR evolves and becomes more complex (as it will be shown below). 
 
 This AR was interesting in respect of its magneto-morphological properties. At this point I have to briefly recall essential moments of the newly developed \citep[][and references herein]{Abramenko2021} Crimean magneto-morphological classification (MMC) of ARs. MMC is based on the idea that while a magnetic tube/bundle rises through the convection zone (CZ), it can undergo certain distortion due to turbulence in the CZ. Flux tubes with low distortion will appear on the surface as regular bipolar magnetic structures following the essential empirical laws \citep{Lidia2015}, that agree well with the mean field dynamo theory (Hale polarity law, Joy's law, dominance of the leading spot). Such ARs (with minimal influence of CZ-turbulence) belong to the MMC-class A and referred to as regular ones. When a bipolar AR violates at least one of the aforementioned laws (MMC-class B1), it can be considered as a result of mild distortion of a single toroidal flux tube. B2 - multipolar ARs consisting of several quasi-coaligned bipoles having general axis orientation in accordance with Joy's law. B2-class ARs may be regarded as the result of fragmentation and distortion of a single toroidal flux tube. B3 - multipolar ARs where opposite polarity sunspots are distributed in an irregular manner, where several flux tubes emerged  independently; these ARs represent the most complex magnetic structures and can be considered as a result of interaction (intertwining) of several flux tubes in the CZ. B1-, B2-, and B3-class ARs are joined into the class B of irregular ARs. Unipolar sunspots are gathered into class U. Thus, the MMC approach allows to reveal an influence of the CZ-turbulence on the formation of active regions.

 According to MMC, the considered AR NOAA 13354 was an irregular one, i.e.,  belonging to class B. The AR emerged on 26-27 of June as a "wrong" bipole B1 (the tailing spots exceeded the leading ones, see the left top panel in Fig. \ref{fig1}); next three days it was classified as B2-class AR - a multipolar structure consisting of several coaligneg bipols (two leading spots and two trailing ones, see the left middle panel in Fig. \ref{fig1}), that can be treated as a result of sub-photospheric splitting of a single flux tube. Finaly, since 1st of July it became clear that the emerged $\delta$-structure was developing into a mature additional bipolar structure implying that at least two separate flux tubes emerged in the place, which corresponds to the B3-class AR. Timing of the transitions B1-B2-B3 is shown in Figures \ref{fig3} and \ref{fig5}.
 So, during 5 days the AR 13354 evolved from an irregular bipole B1 to a complicated B3-class multipolar region. This implies that the original intertwined magnetic knot was formed beneath the photosphere (due to turbulent influence of the CZ) and appeared on the surface gradually, as a sequence of separate chaotically distributed magnetic elements. It was interesting to explore how the turbulence and multifractality parameters varied along such a transformation.

We used magnetograms acquired with the {\it Helioseismic and Magnetic Imager} (HMI) on board the {\it Solar Dynamics Observatory} (SDO) \cite{Scherrer2012}, \cite{Schou2012}. 
 Unlike the data used in the previous study \citep{Abramenko2005a}, when high-resolution magnetograms for an AR were available for several hours only, the HMI data will allow us to explore the time variations of turbulence and multifractality during at least a 5-day-interval centered at the moment of the AR's crossing the central meridian.
To retrieve the data, we downloaded {\it Space-weather HMI Active Region Patches} (SHARP, hmi.sharp\_cea\_720s...magnetogram series) from the {\it Joint Science Operations Center} (JSOC, \url{http://jsoc.stanford.edu/}). The magnetograms were acquired in the Fe I 6173.3\AA\ spectral line with the spatial resolution of 1~arcsec and cadence of 12 minutes. 
Note that sharp\_cea\_720s data provide the line-of-sight (LOS) component of the magnetic field remapped on the grid corrected for the projection effect \citep{Bobra2014}. According to \citet{Liu2012} the standard deviation noise level of this data is about 6.3 Mx cm$^{-2}$. Three magnetograms of the active region are shown in Figure \ref{fig1}, left column: during emergence (27.06.2023/00:36~UT, top), at the moment of the M3.8 flare onset (29.06.2023/14:12~UT), and during the emergence of the $\delta$-structure (30.06.2023/19:48~UT).  

To visualize the M3.8 flare, we used the SDO  {\it Atmospheric Imaging Assembly} (AIA) data \citep{Lemen2012} in the 1600\AA\ spectral line (see the right column frames in Figure \ref{fig1}). 
    
 \section{Turbulence via magnetic power spectra}
 \label{S-Ek} 
 
 A straightforward way to monitor the turbulence state is to explore the time variations of the power spectrum of the field under study. A system in a state of fully developed classical (Kolmogorov) turbulence displays the power spectrum, which follows the power law with the slope (spectral index) of -5/3 \citep{Kolmogorov1941}. Here we used the earlier developed code \citep{Abramenko2001PS,Abramenko2005PS} to calculate the one-dimensional power spectrum of a two-dimensional structure, a magnetogram. The code calculates the squared Fourier transform of the source 2D array and then integrates the energy inside thin annuluses in the  2D wavenumber space. The resulting spectra in the double-logarithmic plot for four magnetograms are shown in Figure \ref{fig2}.
 
 As it was shown in \citet{Abramenko2001PS}, a resolution of an instrument restricts the high-wavenumber boundary of the interval (the inertial range of turbulence), where the spectral index can be determined reliably. The right dotted vertical segment in Figure \ref{fig2} marks this boundary, which corresponds to the spatial scale of 2.5 Mm for magnetograms with resolution of about 0.5$\times$ 2 arcsec \citep{Abramenko2010PS}. 
 
 The lower wavenumber boundary of the inertial range (the left dotted vertical segment on Figure \ref{fig2}) is not strictly specified, however, analysis of many ARs (see, e.g. \citet{Abramenko2005PS,Abramenko2010PS}) demonstrates that on spatial scales above 10~Mm the influence of large sunspots of the AR (that definitely cannot be attributed to turbulence) starts to distort the turbulence spectrum. However, this does not occur in every case. Anyway, for the most reliable determination of the power index it is best to take the scale interval as (2.5 - 10) Mm, or, in the wavenumber scale, (0.628 - 2.51) Mm$^{-1}$.
  
 \textsc{\begin{figure}
 	\centerline{\includegraphics[width=1.0\textwidth,clip=]{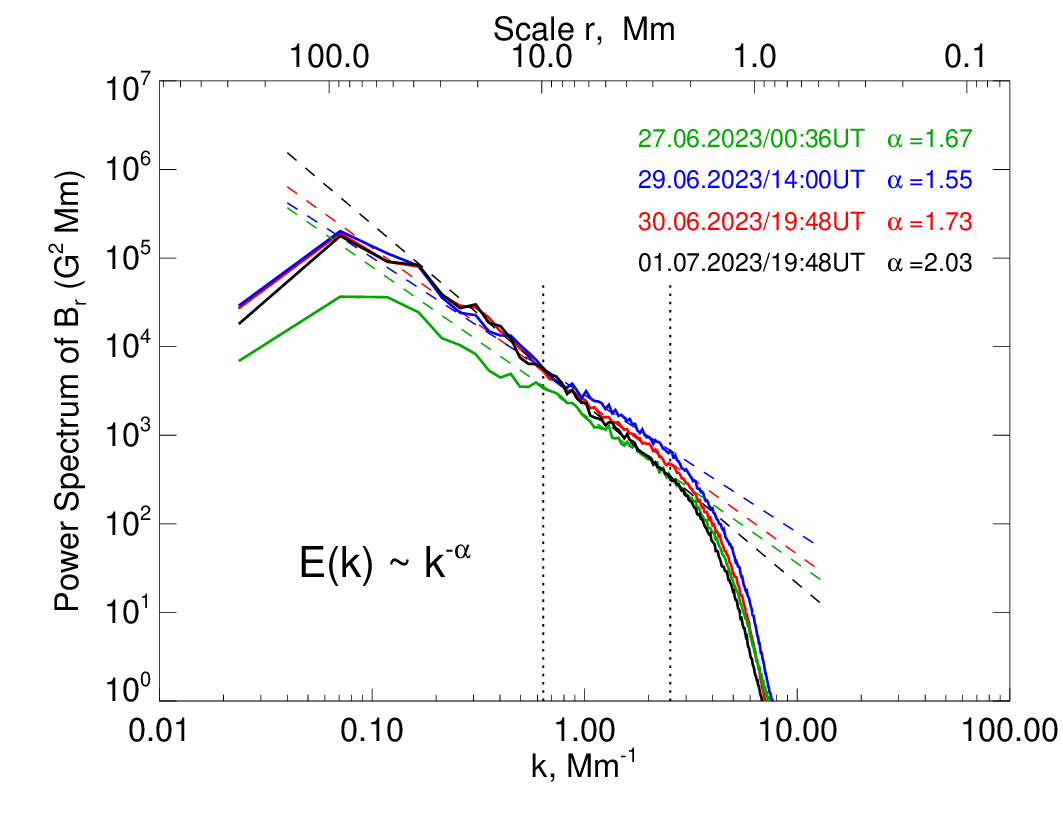}
 	}
 	\caption{ Magnetic power spectra for three magnetograms shown in \ref{fig1}. Green - during the AR's emergence; blue - at the start of the M3.8 flare; red - during the emergence of the $\delta$-structure;  black - at the end of observations, 27 hours prior to the X1.1 flare. Vertical dotted segments mark the inertial interval where the power index $\alpha$ was calculated as a slope of the best linear fit (dashed lines) inside the inertial range of (2.5 - 10) Mm. }
 	\label{fig2}
 \end{figure}}

 \textsc{\begin{figure}
 		\centerline{\includegraphics[width=1.0\textwidth,clip=]{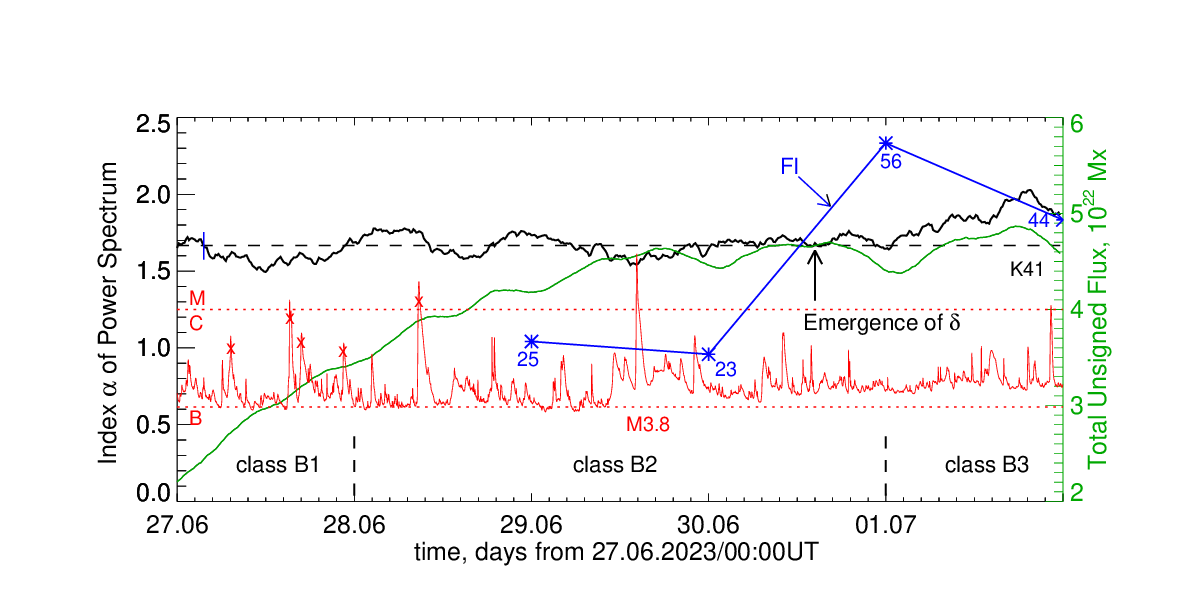}
 		 }
 		\caption{ Time variations of the index $\alpha$ of the magnetic power spectrum (black curve). The blue vertical segment is the typical plus-minus triple standard deviation in calculations of the index as a best linear fit inside the inertial range. Dashed line K41 shows the index $\alpha$=5/3 of Kolmogorov's spectrum of fully developed turbulence. Green curve (right axis) shows the time profile of the total unsigned magnetic flux of the AR. GOES-18 data are shown in red, crosses mark flares launched by other active regions on the disk; M3.8 flare in AR 13354 is marked. Blue stars show the flare index FI calculated as a 4-day running box average by Equation (\ref{FI}). Times of transitions from B1 to B2 to B3 MMC-classes are marked by vertical dashed segments.   }
 		\label{fig3}
 \end{figure}}
 During the emergence phase (green line in Figure \ref{fig2}) the spectrum shows the lower (as compared to the later time) energy at all scales. However, inside the inertial range the slope $\alpha$  of the spectrum $E(k) \sim k^{-\alpha}$ does not vary much until the emergence of the $\delta$-structure. Toward the end of observations, the spectrum (black line in Figure \ref{fig2}) becomes steeper, especially inside the inertial range. This spectrum with the slope of -2.03 was retrieved about 27 hours prior to the X1.1 flare, when the AR was already classified as B3. 
  	
  Figure \ref{fig2} also clearly demonstrates that on scales of approximately 2.5 Mm the spectral cutoff is observed for all spectra. As it was shown in \citet{Abramenko2001PS}, this is the common situation due to influence of the modulation transfer function of the telescope: the artificial cutoff occurs  at the triple resolution limit. Thus, for the MDI full-disk magnetograms, for example, the cutoff is at scales of about 9-10 Mm.
  This is a reason for all MDI-full-disk-spectra in \citet{Georgoulis2012} to be highly non-Kolmogorov ones as soon as the slope was derived from the (3-10) Mm interval.

 Figure \ref{fig3} shows the time variations of the power index $\alpha$ during the 5-day interval. The figure also shows the total unsigned magnetic flux of the AR (green line, right axis) calculated as a sum of absolute values of the magnetic flux density multiplied by the pixel size. The summing was performed over pixels with the flux density over 18 Mx cm$^{-2}$, which corresponds to the three-sigma noise level. The time profile of the total flux demonstrates that the smooth emergence of the AR ceased by noon of 29.06.2023, several hours before M3.8 flare onset. An arrow in Figure \ref{fig3} marks the beginning of emergence of the $\delta$-structure and subsequent formation of a mature bipole to the north from the leading spots.

 Variations in the flaring activity are presented in Figure \ref{fig3} in red. M3.8 flare was launched on 29.06.2023 at 14:00~UT, reached the maximum at 14:15~UT and ended by 14:23~UT. A long period (at least, of two days) of low flaring lasted after this flare. We see that M3.8 flare occurred without any systematic changes in the $\alpha$-profile before or after the flare, amidst the evenly undulating turbulence level. 
 A black dashed line marked as K41 in Figure \ref{fig3} shows the state of classical Kolmogorov turbulence with the power index $\alpha$=5/3. Measured values of $\alpha$ undulate around the K41 line with deviations between 1.5 and 1.8. A rather smooth steepening of the spectrum, up to $\alpha$=2, is observed after the emergence of the $\delta$-structure, i.e., during the formation of a mature bipole. Our monitoring stopped approximately a day before the strong X1.1 flare. Presumably the observed changes in turbulence are the preparation for future flaring, albeit, no clue for the timing prediction. The suggestion is supported by the growth of the flare index by 01.07.2023 (see the blue stars in Figure \ref{fig3}).
 
It is useful to compare the result here with those reported in previous publications. The time variations of $\alpha$ for the mature complex AR NOAA 9077 during two short intervals are shown in Fig. 2 in \citet{Abramenko2005PS}. On those days, the active region was in a well-developed state and can be classified as B3. Smooth variations of $\alpha$ between 2 and 2.5 are observed a day before the strong X-class flare, as well as during the flare. Similar to what is observed in the present study, the flare onset is not accompanied by any abrupt features in the pre-flare $\alpha$-profile, while the general level of $\alpha$ corresponds to that of highly-flaring ARs, see Fig. 2 in \citet{Abramenko2010PS}.  

An interesting case is shown in Figures 3, 4 in  \citet{Abramenko2005PS}: a so called "born bad" active region NOAA 10365, which displayed a steep spectrum, $\alpha$ = (2.1 - 2.2), from the beginning of emergence, throughout the series of M-class flares. Again, we observe no regular specific features in the time profile of $\alpha$ before the flare onsets, except for random undulations. The level of $\alpha$ corresponds to that of highly-flaring ARs, and the MMC-class of this AR can be assigned as B3. It is noteworthy that such "born bad" ARs (with the very steep spectrum from the beginning of emergence and high flaring in future) are rare: a statistical analysis of emerging ARs cannot distinguish them \citep{2021MNRAS.501.6076K}. Some of such ARs are listed in Table 2 in \citet{Abramenko2005PS}.  Figure 3 in the present paper shows that the AR 13354 is not a "born bad" one: the magnetic power spectrum starts with the Kolmogorov-type one and in 5 days smoothly becomes steeper.     

Finaly, the time profile of $\alpha$ for a flare-quiet AR NOAA 10061 (MMC-class A1) is shown in Fig. 7 in \citet{Abramenko2005PS}. A mature stable bipolar flare-less active region during at least 8 hours shows the Kolmogorov-type spectrum with $\alpha$ of (1.55 - 1.77). Of cause, this is a case study, however, the diagram: "Power index $\alpha$ versus Flare index" in Figure 2 in \citet{Abramenko2010PS} shows that a Kolmogorov-type spectrum is not rare for low-flaring ARs: for $\alpha \approx $ 5/3 the flare index varies in a range of (0.05 - 20). For comparison, for high-flaring ARs the flare index can reach up to (200 - 250).  

With the flare index $FI(total)$ = 35, our AR 13354 on the diagram: (Power index $\alpha$ versus Flare index) in Figure 2 in \citet{Abramenko2010PS} shifts from the periphery of the scatter-cloud toward its center, to the regression line. This means that the AR is quite typical and, as such, complies with the inferences of \citet{Abramenko2010PS}  that the turbulence state of the photospheric magnetic field is in a statistical accordance with the flaring productivity of an AR. In spite of that, the present study shows that the timing of flare events is independent from the photospheric turbulence behavior.

 \section{Multi-fractality via flatness function}
\label{S-Fr} 

Another way to challenge the complexity and hierarchical organization of a system is to study the high statistical moments' behaviour. The above discussed power spectrum approach is based on the second statistical moment: using the squared Fourier transform we explore the squared magnetic field density. Using magnetograms of about 10$^{6}$ pixels in size we can explore the statistical moments $q$ up to $q=6$. Multifractality, as a signature of complexity, becomes pronounced in high statistical moments, at least, higher than $q=4$ \citep {Frisch1995}. We have to note here that usage of the structure function for $q=3$ was the main reason for the dissenting inference about multifractality in \citet{Georgoulis2012}.  

We calculate the structure functions $S_q(r)$ of the statistical moments $q=2$ and $q=6$. A ratio of the 6th structure function to the cube of the second structure function gives us the flatness function $F(r)$ with the scaling index $\kappa$ \citep{Frisch1995,Abramenko2005a}: 

\begin{equation} 
F(r)=S_6(r)/(S_2(r))^3 \sim r^{ - \kappa}.
\label{Fr}
\end{equation} 

Here structure functions $S_6(r)$ and $S_2(r)$ are calculated as:

\begin{equation}
S_q(r) = \langle | {B}({\bf x} + {\bf r}) - {B}({\bf x})|^q \rangle, 
\label{Sq} 
\end{equation}  
where ${B}({\bf x})$ denotes the magnetic field strength in the current point $ {\bf x}$ on the magnetogram and  {\bf r} is the separation vector between any two points used to measure the increment; angular brackets denote averaging over the magnetogram. 
For a non-intermittent, mono-fractal structure (e.g., Gaussian field) the flatness function is flat (does not depend on scale); on the contrary, for an intermittent, multi-fractal structure, the flatness grows as a power-law, when the scale $r$ decreases \citep{Frisch1995}. Usually, when studying natural multi-fractals, the range of the flatness’s growth is bounded \citep{Frisch1995}.  
The power index $\kappa$ of flatness function, determined within the power-law range $\Delta r$, is thought to be related to the degree of multifractality: the steeper the function $F(r)$, the higher the degree of multifractality (i.e., the more broad is a set of mono-fractals that contribute into formation of the resulting multifractal, see examples in \citet{Abramenko2005a}).

\begin{figure}
	\centerline{\includegraphics[width=1.0\textwidth,clip=]{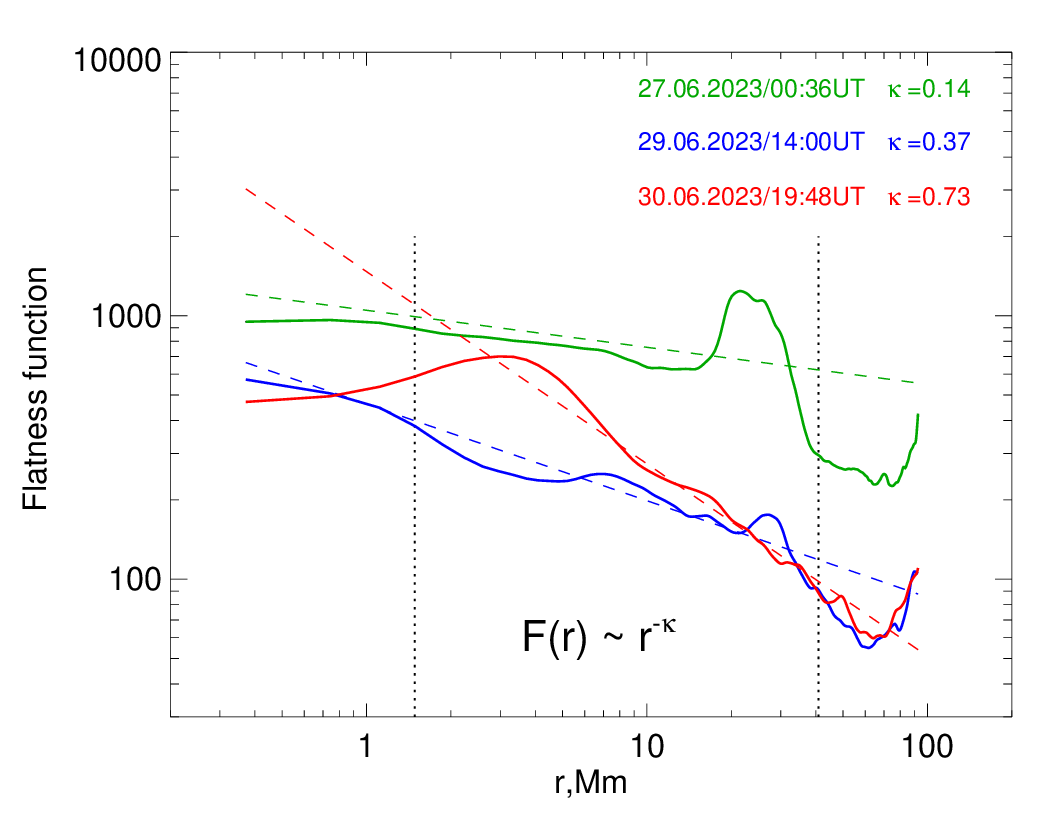}}
	\caption{Flatness functions for three magnetograms shown in \ref{fig1}. Index $\kappa$ for each magnetogram is noted. Notations are the same as in Figure \ref{fig1}. }
	\label{fig4}
\end{figure}

For three magnetograms shown in Figure \ref{fig1} the flatness functions are presented in Figure \ref{fig4}. The interval $\Delta r$ to calculate the power-law index $\kappa$ was selected the same for the entire data set: from 1.5 to 40 Mm. This was done in order to make  the results comparable for all 599 magnetograms. We see that the flatness functions in Figure \ref{fig4} much more differ from each other than the magnetic power spectra do (compare with Figure \ref{fig2}). Indeed, the 6th statistical moment better shows individuality of magnetograms. The scaling indices $\kappa$ differ significantly. One can see that the location of the power-law interval  varies from the emergence stage (green line) to the later times, during the $\delta$-structure appearance (red line): the multifractality regime shifts from small scales (1.5 - 15) Mm toward the larger scales (4 - 70) Mm, and it becomes much stronger.

\begin{figure}
\centerline{\includegraphics[width=1.0\textwidth,clip=]{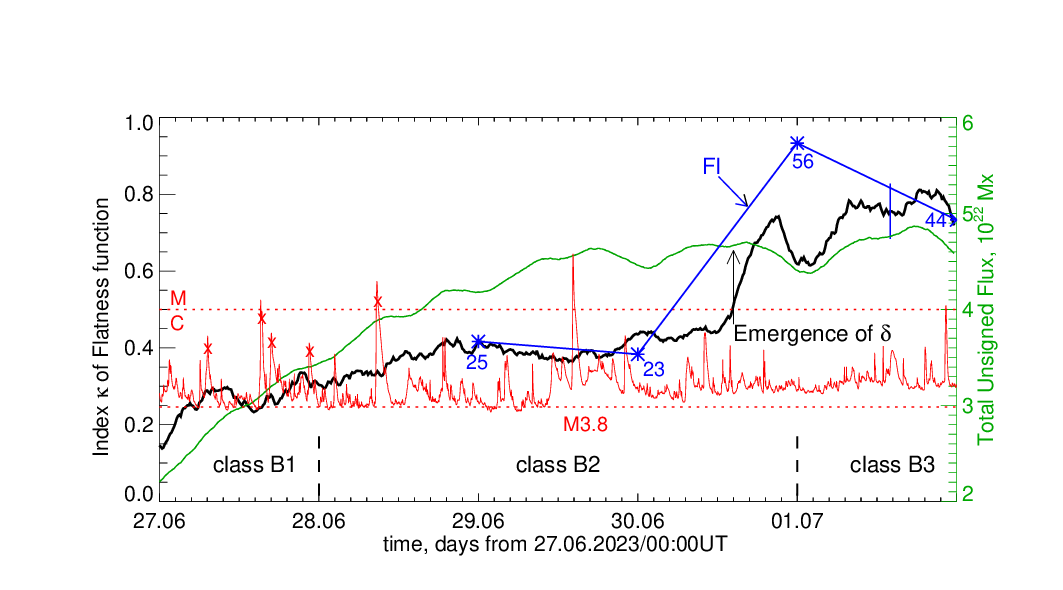}}
\caption{Time variations of the index $\kappa$ of the flatness function (black curve). Notations are the same as in Figure \ref{fig3}.}
\label{fig5}
\end{figure}

Figure \ref{fig5} shows the time variations of the multifractality index $\kappa$ during the 5-day interval of observations of the AR. No noticeable changes in $\kappa$ happened before, during and after the M3.8 flare.  
The only noticeable and reliable variation in the $\kappa$ time profile occurred during the transition to the B3-class (when the $\delta$-structure started to form), two days prior to the X1.1 flare.

The situation suggests that the multifractality index $\kappa$, similar to the turbulence exponent $\alpha$, is prone to enhance with complexity of the magneto-morphological structure and elevation of the flare productivity, however, it does not react to individual flare onset.     

 \section{Concluding remarks}
 \label{S-Con}

The presented analysis of time variations of turbulent and multifractal parameters of the photospheric magnetic field in AR NOAA 13354, along with magneto-morphological transformations and timing of the flaring process in this AR, have shown the following.

Time variations of the magnetic power spectrum exponent $\alpha$ and of the multifractalty exponent $\kappa$ demonstrate no pre-flare or post-flare abrupt peculiarities, instead, 
long periods of stability with smooth transitions into other conditions were observed. 

During the 5-day period studied, the AR emerged and reached the well-developed state, undergoing the magneto-morphological transformations from a single bipole with dominating trailing spot (MMC-class B1) to a configuration of two co-aligned bipoles (MMC-class B2) and to a complex multipolar configuration  of MMC-class B3. The AR launched two strong flares: M3.8 at W12 (this one was in the middle of our interval of monitoring), and an X1.1 flare at W60, which was out of our interval. Both indices, $\alpha$ and $\kappa$, demonstrated no reaction on the first M3.8 flare, neither before, nor after the flare. About 2 days before the X1.1 flare (near the time of the $\delta$-structure appearance and the transition of the AR to B3-class) both indices started a persistent enhancement up to values typical for highly-flaring ARs.

 A conclusion might be inferred that the turbulence and multifractality time path in the photospheric magnetic field does not follow the timing of single flares,  however, it tends to correspond to the  high levels of the magneto-morphological complexity and flaring productivity of an AR.

The result is in a qualitative agreement with a set of previous publications. For example, \citet{Karimov2024} explored the time variations of the free magnetic energy amount in six mature active regions with X-class flares and found that there is no clear correlation between the magnetic free energy and flare peak flux; there is a trend to have strongest flares during the long periods when the free energy exceeds the level of 85\% of its maximum, however, for a given magnitude of the free energy, flares of different classes were observed. As follows from Figures 3 and 4 in \citet{Karimov2024}, the time variations of the free energy amount are rather smooth (especially, for the 2D version of the free energy) and there is no one-to-one correspondence with the flare timing. A similar result was reported by \citet{Fursyak2020}, where for six ARs the time variations of parameters of electric currents were analyzed.  Independent from flaring time, smooth variations of the large-scale global electric current were found for all ARs except for one (NOAA 12192), where the smooth enhancements of the global current were observed in broad time intervals covering a set of strong flares. Again, no one-to-one correspondence with flare onsets were observed. 

Presumably, the photospheric magnetic field and the coronal magnetic field act independently in a sense of criticality \citep[see, e.g.,][]{Dimit2009}. Energy release in flares occurs in the regime of self-organized criticality \citep[SOC,][]{Aschwanden2016,McAteer2016}, when avalanches-flares of all sizes are possible at any time. At the same time, as it was shown recently by \citet{Abramenko2024}, the photospheric magnetic field does not display the property of SOC, its correlation functions follow exponential law, which, according to \citet{Watkins2016,Aschwanden2018}, implies that the field is in a state of self-organization, but without self-organized criticality. 
Note that the idea of self-organization in turbulence in application to solar active regions was brought into attention in the early 2000s, see, e.g., \citet{Vlahos2004}, and later further developed in \citet{Georgoulis2012}.

Meanwhile, for a long time it is known that strongest flares tend to appear in the most complex mature active regions \citep[][to mention a few]{Priest1984,Schrijver2009,Abramenko2021}. As it was mentioned in the Introduction, there are also publications \citep{Hewett2008,McAteer2010,Abramenko2010PS,Abramenko2010Int} stating that there is some positive correlation between the scaling indices (as complexity measures) and the flare productivity.
 The present analysis shows that the future flare productivity is rather determined by large-scale magneto-morphological environment of an AR, formed deep beneath the photosphere long before the appearance of the entire stored flux on the surface. The complexity of this flux determines the future capability for flaring in the corona, and the scaling indices of the field in the photosphere only passively reflect the overall level of complexity, independent of the flare timing. At this point I agree with a statement suggested by \citet{Georgoulis2012} that morphological and topological properties of an AR might be very useful for flare prediction. This statement was confirmed recently by \citet{Abramenko2021}: 63\% out of all ARs of the 23rd and 24th solar cycles that produced X-class flares were classified as B2- and B3-class ARs.

All these together allows us to conclude that the photospheric magnetic field in an AR develops in a state of self-organization, in a state of quasi-stationary regime of fully developed turbulence, but magnetic complexity is predetermined, so that it ensures the observed flaring productivity. Presumably, the fate of an active region is predestined by the sub-photospheric convection long before the appearance of the AR on the solar surface.

\section{Acknowledgments}
Author is thankful to anonymous referee for very important remarks and suggestions. 
SDO is a mission for NASA Living With a Star (LWS) program. The SDO/HMI data were provided by the Joint Science Operation Center (JSOC).

\begin{fundinginformation}
	No funding
\end{fundinginformation}

\begin{dataavailability}
	The SDO/HMI data are available via the {\it Joint Science Operation Center} (JSOC, http://jsoc.stanford.edu). The data obtained in the paper can be offered by the author by request.  
\end{dataavailability}

\begin{ethics}
	\begin{conflict}
		The author declares no conflict of interests.
	\end{conflict}
\end{ethics}

% format of references provided by the journal (.bst)
\bibliographystyle{spr-mp-sola}
% name your Bibtex file containing your references (.bib)
\bibliography{abramenko}  

% Checking: look if the file containing the ``\bibitem'' exits
%           so check if the .bbl file exist (bibTeX compilation)
\IfFileExists{\jobname.bbl}{} {\typeout{}
	\typeout{****************************************************}
	\typeout{****************************************************}
	\typeout{** Please run "bibtex \jobname" to obtain} \typeout{**
		the bibliography and then re-run LaTeX} \typeout{** twice to fix
		the references !}
	\typeout{****************************************************}
	\typeout{****************************************************}
	\typeout{}}

\end{article} 

\end{document}